\documentclass[11pt]{article}
\usepackage{amsmath,amssymb,epsfig,latexsym}

\newtheorem{lemm}{Lemma}

\makeatletter\@addtoreset{equation}{section}\makeatother
\begin{document}
\title{Estimation procedures for
a semiparametric family of bivariate copulas}
\author{C\'ecile Amblard$^{1}$ \& St\'ephane Girard$^{2}$}
\date{
\small
$^{1}$ LabSAD, Universit\'e Grenoble 2,
BP 47, 38040 Grenoble Cedex 9, France.\\
T\'el : (33) 4 76 82 58 26,  Fax : (33) 4 76 82 56 65,
E-mail : Cecile.Amblard@upmf-grenoble.fr\\
$^{2}$ SMS/LMC, Universit\'e Grenoble 1,
BP 53,  38041 Grenoble Cedex 9, France.\\
T\'el : (33) 4 76 51 45 53, E-mail: Stephane.Girard@imag.fr\\
}
\maketitle

\begin{abstract}
\noindent In this paper, we propose simple estimation methods dedicated
to a semiparametric family of bivariate copulas.
These copulas can be simply estimated through the estimation 
of their univariate generating function.  
We take profit of this result to estimate the associated
measures of association as well as the high probability regions
of the copula. These procedures are illustrated on simulations
and on real data.  \\



\end{abstract}

\newpage
                   \section{Introduction}
The theory of copulas provides a relevant tool to build multivariate probability laws, from fixed margins and required degree of dependence. From Sklar's Theorem~\cite{SKLAR}, the dependence properties of a continuous multivariate distribution $H$ can be entirely summarized, independently of its margins, by a copula, uniquely associated with $H$. Several families of copulas, such as Archimedian copulas~\cite{GENEST} or copulas with polynomial sections~\cite{QUESADA,NELSEN97} have been proposed. 
More recently, we proposed to give up the polynomial form to work with a semiparametric family of copulas~\cite{Nous}.
This permits to increase the dependence degree and to preserve the
dependence properties of copulas with polynomial sections 
without significantly complexifying the model.
Furthermore, the family of copulas is generated as simply as Archimedian
copulas, that is by an univariate function.\\
Among the numerous papers dedicated to the construction of copulas, very few
of them propose some associated inference procedures. Besides, starting
from data, it is very difficult to find "Which copula is the right one?",
see~\cite{GRO} for a review on this problem in the financial modeling context.
The first attempt to estimate copulas is achieved in~\cite{DEH} with
the introduction of nonparametric estimates based on empirical copulas.
A recent contribution to the nonparametric estimation of copula for time series is presented in~\cite{SCI1}.
An alternative approach is to choose a parametric family of copulas
and to estimate the parameters by a maximum likelihood method~\cite{JOE2}.
Both approaches suffer from their own drawbacks. 
In the first case, a fully nonparametric estimate is likely to 
suffer from the so-called ``curse of dimensionality''. It would
have a high variance for large number of margins and moderate size of datasets.
At the opposite, a parametric estimate can lead to a very high bias
if the prior family of copulas is not appropriate.
Restricting to Archimedian copulas, Genest \& Rivest~\cite{GENEST2} proposed
a semiparametric estimate. In this family, estimating
the copula reduces to estimating the generating function.
It is therefore realistic to consider a nonparametric
estimate of this univariate function. Here, we show that, similarly,
the estimation of a copula in the semiparametric family~\cite{Nous}
can be simply achieved by estimating nonparametrically the univariate
generating function. This permits to overcome the curse of
dimensionality. We deduce of this estimator some statistical
procedures for estimating the associated dependence coefficients and
high probability regions.\\
In section~\ref{secdef}, the expression of the considered semiparametric family
of copulas is given and its basic properties are recalled.
Section~\ref{estiphi} is dedicated to the estimation of the generating function
and of the dependence coefficients.
In section~\ref{estiregions}, we address the problem of estimating
high probability regions. 
All the proposed estimates are experimented on simulated samples 
in section~\ref{simul} and on real data in section~\ref{realdata}.

            \section{Definition and basic properties}
\label{secdef}

\noindent Throughout this paper, we note $I=[0,1]$. A bivariate copula defined on the unit square $I^2$ is a bivariate cumulative distribution function with univariate uniform $I$ margins. Equivalently, it must satisfy the following properties :
\begin{description}
\item [(P1)] $C(u,0)=C(0,v)=0$, $\forall (u,v)\in I^2$,
\item [(P2)] $C(u,1)=u$ and $C(1,v)=v$, $\forall (u,v)\in I^2$,
\item [(P3)] $\Delta(u_1, u_2, v_1, v_2)=C(u_2,v_2)-C(u_2,v_1)-C(u_1,v_2)+C(u_1,v_1)\geq 0$, $\forall (u_1, u_2, v_1, v_2)\in I^4$, such that $u_1\leq u_2$ and $v_1\leq v_2$.
\end{description}
Let us recall that, from Sklar's Theorem~\cite{SKLAR}, any 
bivariate  distribution with cumulative distribution function $H(x,y)$
and marginal cumulative distribution functions $F(x)$ and $G(y)$ can be written
$H(x,y)=C(F(x),G(y))$, where $C$ is a copula. This result justifies the use of copulas for building bivariate distributions.\\
\noindent Here, we consider the semiparametric family of functions defined on $I^2$ by:
\begin{equation}
\label{defcop}
C_{\theta,\phi}(u,v)=uv+\theta \phi(u)\phi(v), \;\;\theta\in[-1,1],
\end{equation}
where $\phi$ is a function on $I$. This family was first introduced in~\cite{LLALENA}, chapter 3, and is a particular case of Farlie's family~\cite{FAR}. It is extensively studied in~\cite{Nous}. In particular, the following
basic lemma is proved:
\begin{lemm}
\label{lemprinc}
$\phi$ generates a parametric family of copulas
 $\{C_{\theta,\phi},\; \theta\in[-1,1]\}$
 if and only if it satisfies the following conditions~:
\begin{description}
\item [(i)] $\phi(0)=\phi(1)=0$,
\item [(ii)] $|\phi(x)-\phi(y)|\leq|x-y|$ for all $(x,y)\in I^2$.  
\end{description}
\end{lemm}
\noindent
The function $\phi$ plays a role similar to the generating function 
in Archimedian copulas~\cite{GENEST}. Each copula $C_{\theta,\phi}$
is entirely described by the univariate function $\phi$ and the parameter $\theta$, which tunes the dependence between the margins.
For instance, $\phi(x)=x(1-x)$ generates the Farlie-Gumbel-Morgenstern (FGM) family of copulas~\cite{FGM}, which contains all copulas with both horizontal and vertical quadratic sections~\cite{QUESADA}. 
Another example is $\phi(x)=x(1-x)(1-2x)$ which defines the parametric family 
of symmetric copulas with cubic sections proposed in~\cite{NELSEN97},
equation~(4.4).
Of course, it is also possible to choose $\phi$ so as to define new copulas, 
see section~\ref{simul} for an example.

	\section{Estimation of the generating function}
\label{estiphi}

\subsection{Preliminaries}

Let $(X,Y)$ be a random pair from the cumulative
distribution function $H(x,y)=C_{\theta,\phi}(F(x),G(y))$,
where $F(x)$ and $G(y)$ are respectively the cumulative distribution
functions of $X$ and $Y$. 
The estimation of the copula reduces to estimating the generating function
$\phi$ and the parameter $\theta$. This estimation clearly suffers from an
identifiability problem since, for instance,
replacing $\phi$ by $\phi/\sqrt{\alpha}$ and $\theta$ by $\alpha\theta$ for 
any positive $\alpha$ leads to the same copula. 
When $\theta\neq 0$, introducing $s=\theta/|\theta|$ and
$\psi=\sqrt{|\theta|}\phi$ yields
\begin{equation}
\label{newcopu}
C_{\theta,\phi}(u,v)=C_{s,\psi}(u,v)=uv+s\psi(u)\psi(v),
\end{equation}
where $\psi$ satisfies the conditions of Lemma~\ref{lemprinc} and
$s\in\{-1,1\}$.  The identifiability problem is not fully overcome
with the new parameterization~(\ref{newcopu}) since the sign of $\psi$
cannot be identified.  Thus, we limit ourselves to the
Positively Quadrant Dependent (PDQ) context.  
Recall that  $X$ and $Y$ are PQD~\cite{JOE}, section~2.1.1, if and only if
$$
\forall (x,y)\in \mathbb{R}^2,\;\; P(X\leq x,Y\leq y)\geq P(X\leq x)P(Y\leq y).
$$
As shown in \cite{Nous}, theorem~3, $X$ and $Y$ are PQD if and only if 
\begin{equation}
\label{PQD}
\theta>0 \mbox{ and, either }\forall u \in I,\; \phi(u)\geq 0  \mbox{ or }
\forall u \in I,\; \phi(u)\leq 0.
\end{equation}
If $(X,Y)$ are PQD, then the  copula (\ref{newcopu}) can always be rewritten as:
\begin{equation}
\label{fincopu}
C_{1,\psi}(u,v)=uv+\psi(u)\psi(v),
\end{equation}
where $\psi$ is a non negative function satisfying the conditions
of Lemma~\ref{lemprinc}. In the following, we limit ourselves to
the estimation of $C_{1,\psi}$, or equivalently to the estimation
of $\psi$ in this context. We refer to section~\ref{extension} for
possible improvements of the estimation method.  

\subsection{Estimation of $\psi(w)$}

Let $\{(x_i,y_i),i=1,\dots,n\}$ a sample of $(X,Y)$ from the cumulative
distribution function $H(x,y)$.  The rank transformations
$u_i=\mbox{Rank}(x_i)/n$ and $v_i=\mbox{Rank}(y_i)/n$ yield
an approximate sample from the copula $C_{1,\psi}(u,v)$.  
The estimation of $\psi(w)$ relies on the pseudo-observations 
$w_i=\max(u_i,v_i)$, $i=1,\dots,n$
which have the common distribution function
$C_{1,\psi}(w,w)=w^2+\psi^2(w)$.
In order to obtain a regular estimated function, this estimate 
is written as a linear combination of a denombrable
set ${\cal A}$ of functions:
\begin{equation}
\label{expanphi}
\widehat\psi(w)=\sum_{k\in{\cal A}} a_k e_k(w).
\end{equation}
The set of functions $\{e_k(w),\; w\in I,\; k\in{\cal A}\}$
need not to be orthogonal but the condition $e_k(0)=e_k(1)=0$
is required for all $k\in{\cal A}$ in order to ensure 
$\widehat\psi(0)=\widehat\psi(1)=0$,
and thus to respect condition (i) of Lemma~\ref{lemprinc}.
Introducing $w_{1,n}\leq\dots\leq w_{n,n}$
the ordered statistics associated to the $w_i$, $i=1,\dots,n$, 
the coefficients $a_k$ are determined by 
the constrained least-square problem
\begin{equation}
\label{pbopti}
\hat{a}=\arg\min \left\{\|Ma-b\|^2,\; 0\leq (Ma)_i,\; -1\leq (M'a)_i\leq 1,
\; i=1,\dots,n \right\}
\end{equation}
where $M$ and $M'$ are two matrices such that
$M_{i,k}=e_k(w_{i,n})$ , $M'_{i,k}=e'_k(w_{i,n})$
for $k\in{\cal A}$, $i\in\{1,\dots,n\}$,
and $b$ is a vector defined by $b_i=(i/(n+1)-w_{i,n}^2)^{1/2}$,
$i\in\{1,\dots,n\}$.
The minimization of $\|Ma-b\|^2$ ensures that for all $i=1,\dots,n$
\begin{equation}
\label{approxpsi}
\widehat\psi(w_{i,n})^2  = C_{1,\psi}(w_{i,n},w_{i,n})-w_{i,n}^2
\approx i/(n+1) - w_{i,n}^2.
\end{equation}
The positivity conditions $0\leq(Ma)_i$ impose that
$0\leq \widehat\psi(w_{i,n})$, and the bound conditions\\
$-1\leq (M'a)_i\leq 1$
are interpreted as $|\widehat{\psi}'(w_{i,n})|\leq 1$ which allows
to fulfil Lemma~\ref{lemprinc}(ii).\\
In practice, the optimization process provides reasonably sparse solutions,
that is vectors $a$ with a moderate number of nonzero components.  

\subsection{Estimation of an association measure}

Two measures of association between the components of the random pair $(X,Y)$
are usually considered~\cite{JOE}, section~2.1.9.
The Kendall's Tau is the probability of concordance minus the probability of
discordance of two random pairs $(X_1,Y_1)$ and $(X_2,Y_2)$ described by the
same joint bivariate law $H(x,y)=C(F(x),G(y))$. It only depends on the
copula:
\begin{equation}
\label{deftau}
\tau=4\int_{I^2} C(u,v) dC(u,v)\!-\!1.
\end{equation}
The Spearman's Rho is the probability of concordance minus the probability of discordance of two pairs $(X_1,Y_1)$ and $(X_2,Y_2)$ with respective joint cumulative law $H(x,y)$ and $F(x)G(y)$, 
$$
\rho=12\int_{I^2}  C(u,v) dudv\!-\!3.
$$
In~\cite{Nous}, proposition~1, it is shown that, within the
family~(\ref{defcop}),
these two measures are equivalent up to a scale factor: $2\rho=3\tau$.  
Here, we focus on the Spearman's Rho whose expression is in our family:
$$
\rho=12\theta\left(\int_I\phi(u)du\right)^2=12\left(\int_I\psi(u)du\right)^2.
$$
We then propose an estimate based on the estimate $\widehat\psi$:
\begin{equation}
\label{estirho}
\hat{\rho}_{\mbox{\tiny SP}}
 =
12\left(\sum_{k\in{\cal A}} a_k \beta_k \right)^2,
\end{equation}
where we have introduced $\beta_k=\int_I e_k(u)du$. This integral can be 
calculated analytically or numerically by Simpson's rule, depending
on the complexity of the basis of functions.  
It is also possible to estimate $\rho$ in a nonparametric way by rescaling the
empirical version of~(\ref{deftau}) introduced in~\cite{GENEST2}
with the factor $3/2$ to obtain:
$$
\hat{\rho}_{\mbox{\tiny NP}}
=
\frac{6}{n(n-1)}\sum_{i=1}^n\sum_{j=1}^n
{\mathbf 1}\{u_j<u_i,\;v_j<v_i\}- \frac{3}{2},
$$
where ${\mathbf 1}\{.\}$ is the indicator function.
The two estimates $\hat{\rho}_{\mbox{\tiny SP}}$ and
$\hat{\rho}_{\mbox{\tiny NP}}$ are compared
on simulations in section~\ref{simul}.

        \section{Estimation of high probability regions}
\label{estiregions}

\subsection{The general problem}

Let us recall the definition of a $p$-dimensional $\alpha$-quantile 
of a distribution $P$. Let ${\cal S}$ the class of Borel measurable sets
of ${\mathbb R}^p$ and let $\lambda$ be the Lebesgue measure
defined on ${\cal S}$:  
$$
Q_\alpha=\inf\{\lambda(S): P(S)\geq \alpha, \; S\in{\cal S}\},\; 0<\alpha\leq 1.
$$
Here, $Q_\alpha$ is the minimum volume $S\in{\cal S}$ that contains
at least a fraction $\alpha$ of the probability mass.  
This is a particular case of the general quantile function
introduced by Einmal and Mason~\cite{EM}.  
A particular attention has been paid to the estimation of
$Q_1$, the support of the distribution, from a sample
$\{M_1,\dots,M_n\}$ of ${\mathbb R}^p$.  
The early paper of Geffroy~\cite{Geff1} takes place in the case $p=2$ and the
considered supports are written
$$
 Q_1=\{(x,y)\in {\mathbb R}^2: 0\leq x\leq 1~;\;0\leq y \leq f(x)\},
$$
where $f$ is an unknown function. More recently, smooth estimates of
the frontier function $f$ have been proposed~\cite{GirJac} and their
extension to star-shaped supports has been studied~\cite{JacSuq}.
Numerous works have been dedicated to the nonparametric estimate
$$
\widehat{Q}_1 = \bigcup_{i=1}^n B(M_i, r_n),
$$
where $B(M_i,r_n)$ is the ball of radius $r_n$ and centered at $M_i$,
see for instance~\cite{DevWis,Gensb}.
In the latter paper, a partition based estimate is also proposed.  
Introducing $\{K_i\}_{i\geq 1}$ a partition of ${\mathbb R}^2$,
the estimate is simply:
$$
\widehat{Q}_1 = \bigcup_{i,j} K_i {\mathbf 1}\{M_j \in K_i\}.
$$
In subsection~\ref{algo}, we propose an estimate of $Q_\alpha$, $\alpha<1$,
when $P$ is a bivariate copula, with a similar principle.  
Subsection~\ref{estiproba} examines the special case of the 
semiparametric family of copulas $\{C_{1,\psi}\}$.  
The estimation of such high probability regions is of particular
interest for bivariate copulas since some dependence properties can
be read on this graph, see~\cite{LONG}. To this end, introduce
the two diagonal lines $\ell_1$: $v=u$ and $\ell_2$: $v=1-u$ of the
unit square $I^2$.
If the graph of $Q_\alpha$ is nearly symmetric with respect to both
diagonals $\ell_1$ and $\ell_2$ then the copula models weakly dependent
variates. On the contrary, if the graph is concentrated along $\ell_1$, then
the copula models strongly positively dependent variates.

\subsection{The algorithm} 
\label{algo}

\noindent Let $\{I_k,\;k=1,\dots,N\}$ be the equidistant $N$-partition of $I$
and $K_{k,\ell}= I_k\times I_\ell$ the associated $N\times N$ grid.  
Denote $\delta_{k,\ell}\in\{0,1\}$, $k=1,\dots,N$, $\ell=1,\dots,N$
a binary $N\times N$ array and introduce the estimate
\begin{equation}
\label{estiQ}
\widehat{Q}_\alpha = \bigcup_{k,\ell} K_{k,\ell}
{\mathbf 1}\{\delta_{k,\ell}=1\},
\end{equation}
where the $\delta_{k,\ell}$ are defined by the optimization problem
\begin{equation}
\label{objectif}
\min \frac{1}{N^2} \sum_{k=1}^N \sum_{\ell=1}^N \delta_{k,\ell},
\end{equation}
under the constraints $\delta_{k,\ell}\in\{0,1\}$ and
\begin{equation}
\label{contrainte}
\sum_{k=1}^N \sum_{\ell=1}^N \delta_{k,\ell} \widehat{P}(K_{k,\ell})\geq \alpha.
\end{equation}
The quantity $\widehat{P}(K_{k,\ell})$ is an estimation of the
probability $P(K_{k,\ell})$.
The quality of the estimate $\widehat{Q}_\alpha$ strongly depends on
the quality of this estimate.
This is discussed in subsection~\ref{estiproba}.
Let us note that~(\ref{objectif}) is equivalent to minimizing 
$\lambda(\widehat{Q}_\alpha)$ and~(\ref{contrainte})
corresponds to the constraint $P(\widehat{Q}_\alpha)\geq\alpha$.  
This optimization problem can be solved with a simple algorithm.
The first step consists to sorting the estimated
probabilities $\widehat{P}(K_{k,\ell})$ in decreasing order to
obtain the sequence $\tilde{P}_\tau$, $\tau=1,\dots,N^2$.
The second step is the computation of
the number of subsets of the partition which are going to be used:
$$
J=\min \left\{j,\;\sum_{\tau=1}^j \tilde{P}_\tau \geq \alpha\right\}.
$$
The last step is the selection of the $J$ first subsets:
$
\delta_{k,\ell}=1 \mbox{ if } 1\leq\tau(k,\ell)\leq J,
$
which leads to the estimate~(\ref{estiQ}).

\subsection{Estimation of $P(K_{k,\ell}) $}
\label{estiproba}

If no information is available on the distribution $P$,
one can use the nonparametric estimate
$$
\widehat{P}_{\mbox{\tiny NP}}(K_{k,\ell})=\frac{1}{n}\sum_{i=1}^n{\mathbf 1}\{M_i\in K_{k,\ell}\}.
$$
Now, restricting ourselves to the family $\{C_{1,\psi}\}$
allows to consider semiparametric estimates,
which are much more accurate than the nonparametric one.
Taking into account of~(\ref{fincopu}), it follows that
\begin{eqnarray*}
P(K_{k,\ell}) &=&  C_{1,\psi}\left(\frac{k}{N},\frac{\ell}{N}\right) 
		- C_{1,\psi}\left(\frac{k-1}{N},\frac{\ell}{N}\right) 
		- C_{1,\psi}\left(\frac{k}{N},\frac{\ell-1}{N}\right) 
		+ C_{1,\psi}\left(\frac{k-1}{N},\frac{\ell-1}{N}\right)\\
&=& \frac{1}{N^2} +
\left(\psi\left(\frac{k}{N}\right)
- \psi\left(\frac{k-1}{N}\right)\right)
\left(\psi\left(\frac{\ell}{N}\right)
- \psi\left(\frac{\ell-1}{N}\right)\right).
\end{eqnarray*}
Thus, basing on section~\ref{estiphi}, the following semiparametric estimate
can be introduced
$$
\widehat{P}_{\mbox{\tiny SP}}(K_{k,\ell})=  \frac{1}{N^2} +
\left({\widehat\psi}\left(\frac{k}{N}\right)
- {\widehat\psi}\left(\frac{k-1}{N}\right)\right)
\left({\widehat\psi}\left(\frac{\ell}{N}\right)
- {\widehat\psi}\left(\frac{\ell-1}{N}\right)\right).
$$
The results using $\widehat{P}_{\mbox{\tiny SP}}$ and 
$\widehat{P}_{\mbox{\tiny NP}}$ are compared on simulations in 
the next section.  

        \section{Simulation results}
\label{simul}

All the numerical experiments of this section have been conducted on the family
of copulas generated by the set of functions
\begin{equation}
\label{exemple}
\forall k\geq 1, \;\;\psi_{k}(x)=1-\left(x^k+(1-x)^k\right)^{1/k},\; x\in I.
\end{equation}
For the sake of simplicity, the resulting
family of copulas will be denoted by $C_k(u,v)=C_{1,\psi_k}(u,v)$.  
This family is interesting since it can represent very
different distributions:
\begin{itemize}
\item When $k=1$, $C_1(x,y)$ is the uniform distribution on 
the unit square $I^2$.  The associated Spearman's Rho
is $\rho_1=0$.
\item Letting $k\to\infty$, we obtain $\psi_k(x)\to \psi_\infty(x)=\min(x,1-x)$
for all $x\in I$.  Thus, it appears that $C_{\infty}(x,y)$
is a mixture of two uniform distributions on the squares $[0,1/2]^2$
and $[1/2,1]^2$ with mixing parameter $1/2$.  
The associated Spearman's Rho
is $\rho_\infty=3/4$, the maximum value in the family~(\ref{defcop}).
\item When $1<k<\infty$, we get a bivariate distribution
``interpolating'' between the two previous one. Up to our knowledge,
it is not possible to calculate $\rho_k$ explicitly.  
\end{itemize}

\subsection{Simulation of data from the copula}

The simulation method described in~\cite{NELSEN99}, p.~36 is used.
First, two independent uniform
samples $u_i$ and $t_i$, $i=1,\dots,n$ are simulated.  
Second, for each $i=1,\dots,n$, $v_i$ is computed such
that 
$$
t_i= \frac{\partial}{\partial u} C_k(u_i,v_i),
$$
by a dichotomy procedure.  
The resulting sample $(u_i,v_i)$, $i=1,\dots,n$ has joint
distribution function $C_k(u,v)$.  

\subsection{Estimation of the generating function}

\begin{figure}[p]
\noindent\centerline{\epsfig{figure=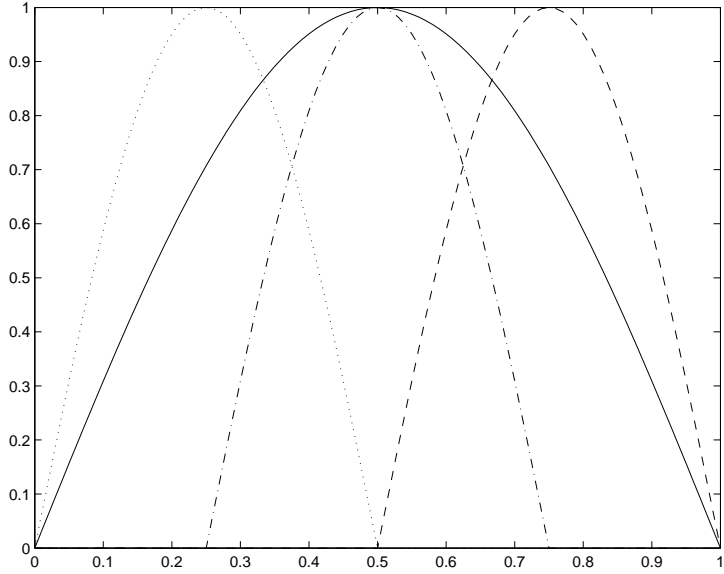,width=0.6\textwidth}}
\caption{Graph of the first basis functions of~(\ref{basetrigo}).
Solid line: $e_{0,0}(x)$, dotted line: $e_{1,0}(x)$, dashdot line: $e_{1,1}(x)$,
dashed line: $e_{1,2}(x)$.  }
\label{dessinbase}
\end{figure}

\begin{figure}[hb]
\noindent\centerline{\epsfig{figure=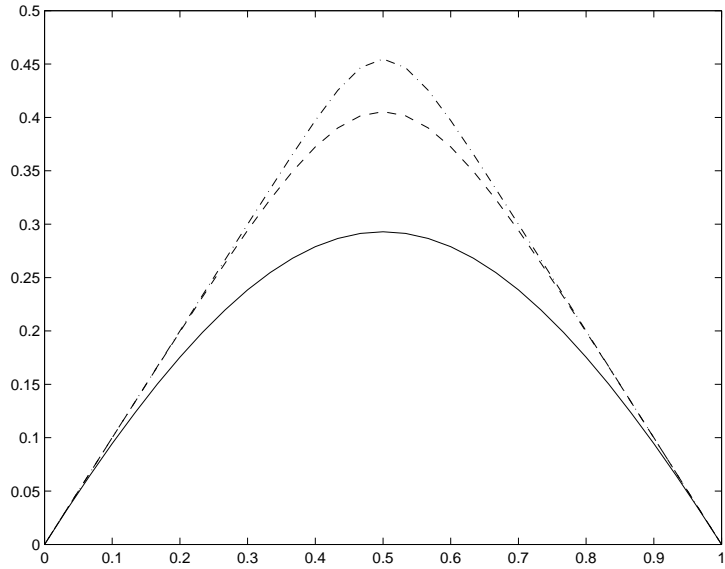,width=0.6\textwidth}}
\centerline{(a) True generating functions $\psi_k(x)$, $k\in\{2,4,8\}$ }
\label{figurephi}

\noindent\centerline{\epsfig{figure=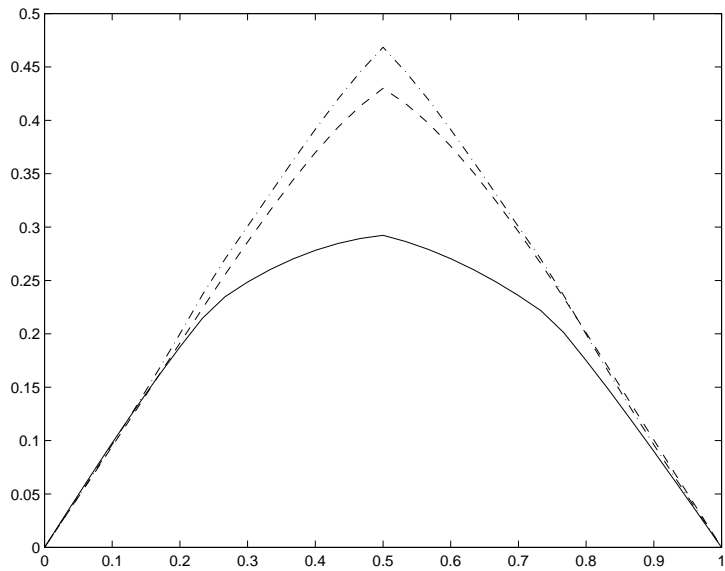,width=0.6\textwidth}}
\centerline{(b) Estimated generating functions $\hat{\psi}_k(x)$, $k\in\{2,4,8\}$ }
\caption{Comparison of the true generating functions $\psi_k(x)$ 
defined by~(\ref{exemple}) and the estimated ones $\hat{\psi}_k(x)$.  
Solid line: $k=2$, dashed line: $k=4$, dashdot line: $k=8$.    }
\label{compaphi}
\end{figure}

In this paragraph, we have chosen $n=100$.  
Starting from the sample $(u_i,v_i)$, $i=1,\dots,n$,
the generating function is estimated using the
procedure described in section~\ref{estiphi}. 
The chosen basis of functions is doubly-indexed by a scale
parameter $s$ and a location parameter $\ell$:
\begin{equation}
\label{basetrigo}
e_{s,\ell}(x)=\sin\left(\frac{\pi}{2}(2^{s+1}x-\ell)\right){\mathbf 1}\{2^{s+1}x\in[\ell,\ell+2]\},
\; (s,\ell)\in{\cal A},
\end{equation}
where ${\cal A}=\{(s,\ell),\;0\leq \ell\leq 2(2^s-1),\; 0\leq s\}$.
See figure~\ref{dessinbase}
for a graph of the first basis functions.  
In figure~\ref{compaphi}, the estimation of $\psi_2(x)$,
$\psi_4(x)$ and $\psi_8(x)$ is compared to the true generating functions.  
These first  results are visually satisfying.  
The optimization procedure selects about 30 basis functions.
A more precise comparison is achieved in table~\ref{tabphi} by
repeating each estimation on 100 different samples.  
On the basis of these 100 repetitions, the mean value and the
standard deviation of the $L_2$ error 
$$
\varepsilon=\left(\int_I (\psi_k(x)-\hat{\psi_k}(x))^2 dx\right)^{1/2}
$$
are evaluated, as well as the mean value and the standard deviation
of the two estimates $\hat{\rho}_{\mbox{\tiny SP}}$ and
$\hat{\rho}_{\mbox{\tiny NP}}$ of the Spearman's rho.  
The integral appearing in $\rho_k$ is
calculated numerically using Simpson's rule.  
The computation of $\hat{\rho}_{\mbox{\tiny SP}}$ is based on~(\ref{estirho})
and is thus explicit after remarking that
$\beta_{s,\ell}=\int_I e_{s,\ell}(x)dx=2^{1-s}/\pi$.\\
The mean values of $\varepsilon$ 
(about $10^{-2}$) confirm that the function $\psi_k$ is correctly estimated. Moreover it appears that the semi parametric estimation of the Spearman's Rho is better than the non parametric estimate, excepted for the case $k=1$. The standard deviations are similar for the two estimates. 

\subsection{Estimation of high probability regions}

\begin{figure}[p]
\noindent\centerline{\epsfig{figure=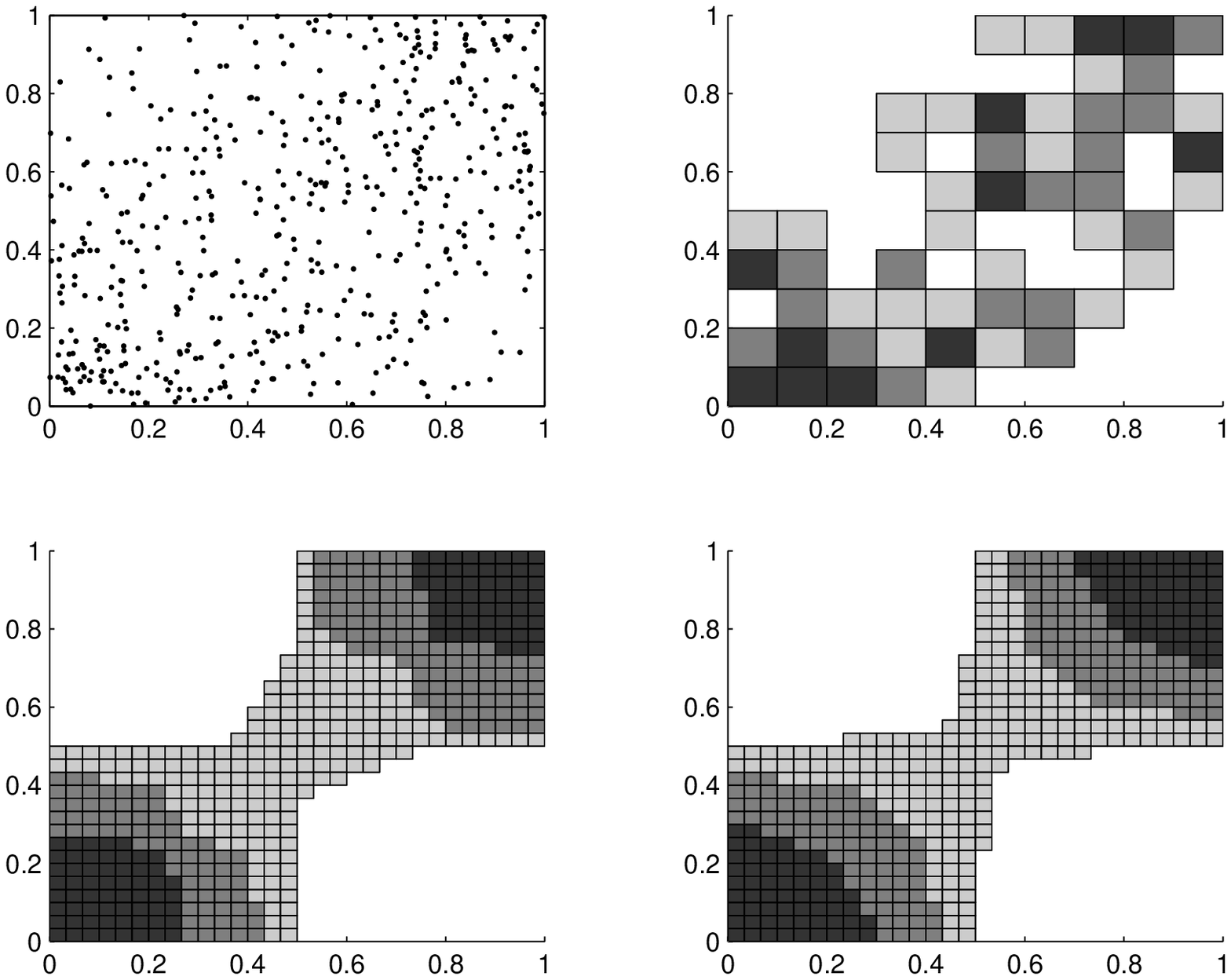,width=0.8\textwidth}}
\caption{Estimation of high probability regions $Q_\alpha$ from $C_2(u,v)$.  
Dark Grey: $\alpha=0.25$, grey: $\alpha=0.5$, light grey: $\alpha=0.75$.  
Top left: simulated sample, top right: nonparametric estimate, 
bottom left: semiparametric estimate, bottom right: semiparametric estimate
with the true function $\psi$.  }
\label{level2}
\end{figure}

\begin{figure}[p]
\noindent\centerline{\epsfig{figure=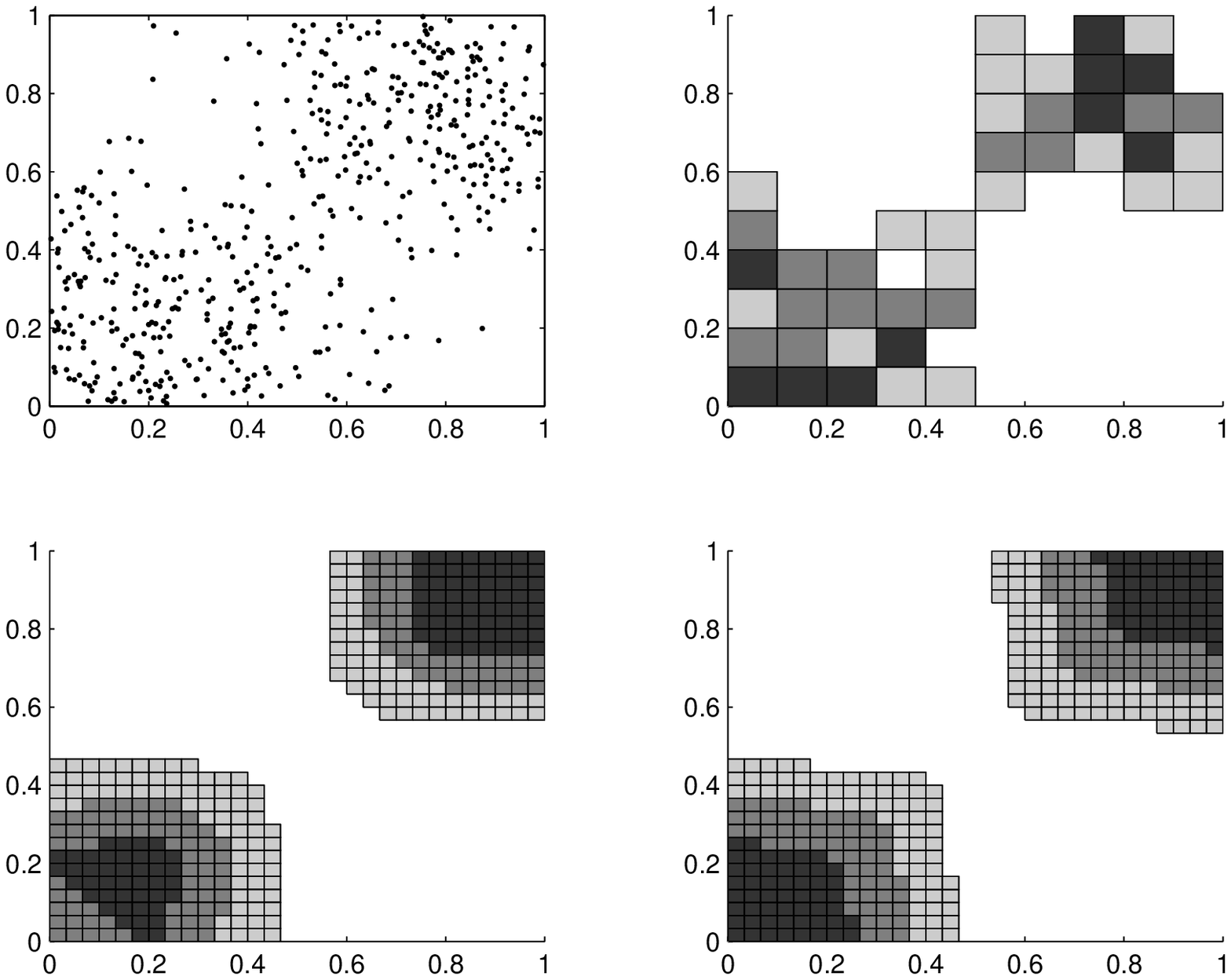,width=0.8\textwidth}}
\caption{Estimation of high probability regions $Q_\alpha$ from $C_4(u,v)$.  
Dark Grey: $\alpha=0.25$, grey: $\alpha=0.5$, light grey: $\alpha=0.75$.  
Top left: simulated sample, top right: nonparametric estimate, 
bottom left: semiparametric estimate, bottom right: semiparametric estimate
with the true function $\psi$.  }
\label{level4}
\end{figure}

\begin{figure}[p]
\noindent\centerline{\epsfig{figure=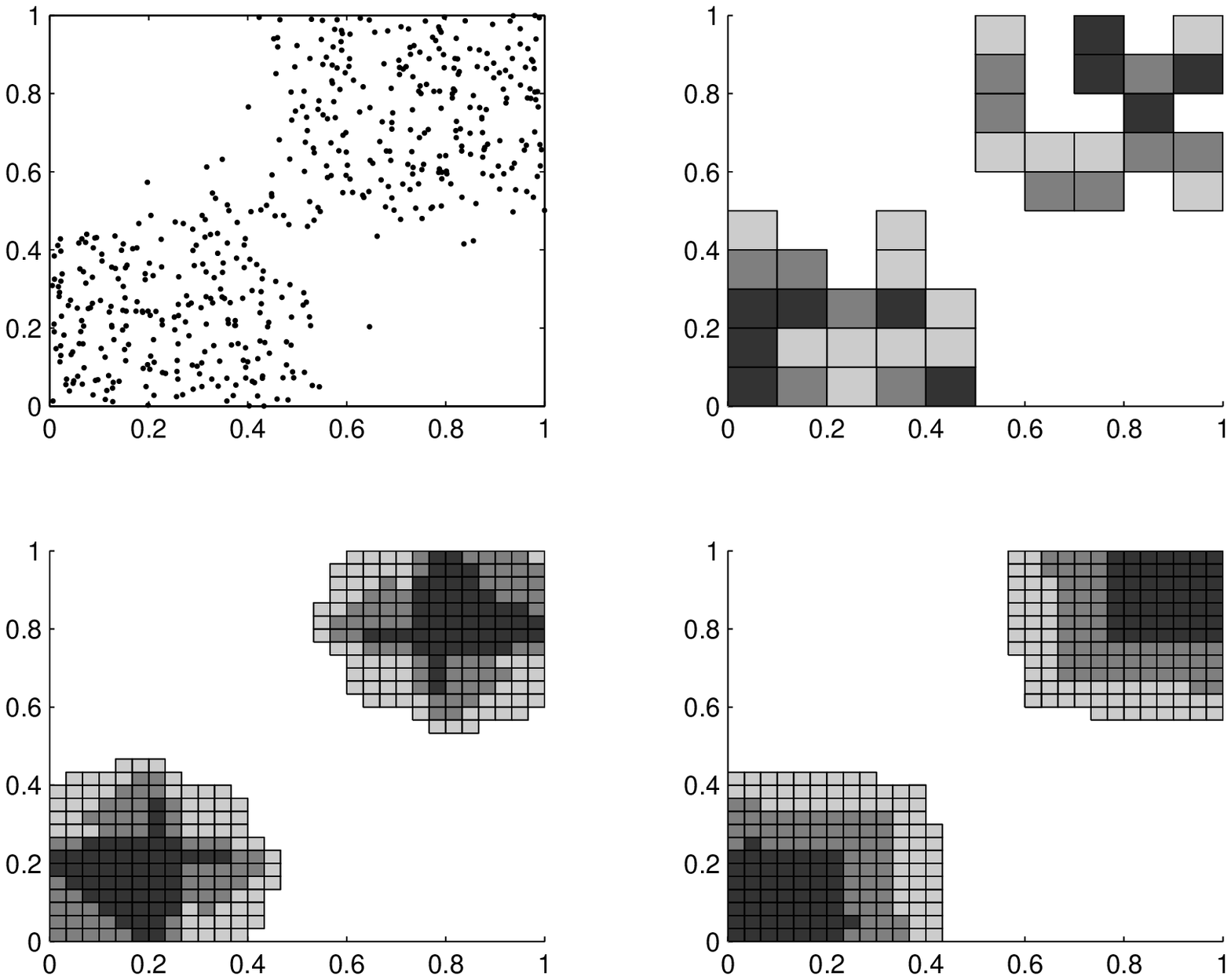,width=0.8\textwidth}}
\caption{Estimation of high probability regions $Q_\alpha$ from $C_8(u,v)$.  
Dark Grey: $\alpha=0.25$, grey: $\alpha=0.5$, light grey: $\alpha=0.75$.  
Top left: simulated sample, top right: nonparametric estimate, 
bottom left: semiparametric estimate, bottom right: semiparametric estimate
with the true function $\psi$.  }
\label{level8}
\end{figure}

Starting from the estimations of $\psi_2(x)$, $\psi_4(x)$ and $\psi_8(x)$
obtained above, it is possible to estimate high probability
regions $Q_\alpha$ with the procedure described in section~\ref{estiregions}.
The following probabilities
$\alpha=0.25$, $\alpha=0.5$ and $\alpha=0.75$ are considered. 
Here, the regions obtained using the semiparametric estimate
$\widehat{P}_{\mbox{\tiny SP}}$, the nonparametric estimate 
$\widehat{P}_{\mbox{\tiny NP}}$ and the true probability
$P$ can be compared.
Of course, this probability depends on $\psi(x)$ and cannot be
used in practical situations.
Here $n=500$ and the estimated regions are obtained by a discretisation on a
$N\times N$ grid with $N=30$ when using $\widehat{P}_{\mbox{\tiny SP}}$
or $P$ and $N=8$ when using $\widehat{P}_{\mbox{\tiny NP}}$ so as
to obtain approximatively $500/64\simeq 8$ points in each cell.
The results are presented on figures~\ref{level2}--\ref{level8}.  
It is apparent that the estimations obtained with the true
probability $P$ and its semiparametric estimate $\widehat{P}_{\mbox{\tiny SP}}$
are very close, especially for moderate values of $k$. 
This confirms the results obtained in the previous paragraph.
On the contrary, the nonparametric estimate accuracy is very poor
for such values of the sample size $n$.
We can also observe that, as $k$ increases, the high probability
regions are more and more concentrated in the neighborhood of
the $\ell_1$ diagonal line. This, together with table~\ref{tabphi},
illustrates the property that the positive dependence is increasing with $k$.

        \section{Real data}
\label{realdata}

\begin{figure}[p]
\noindent\centerline{\epsfig{figure=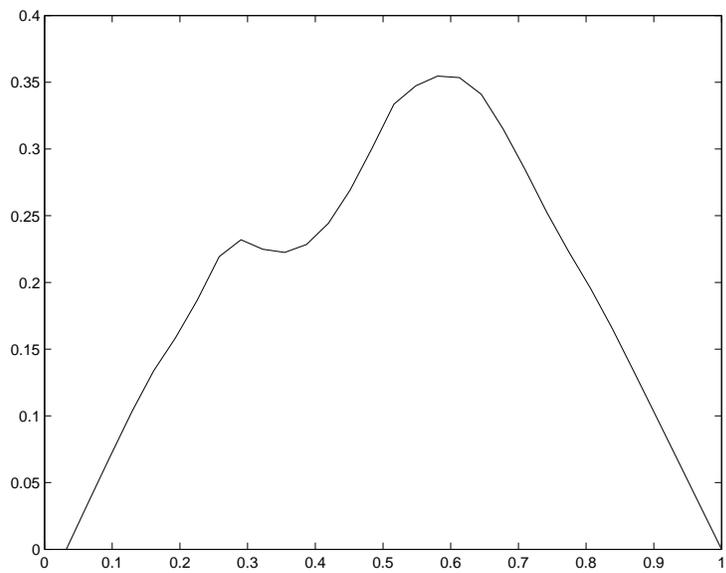,width=0.6\textwidth}}
\caption{Estimation of the generating function $\psi(x)$ from real data.  }
\label{phirealdata}
\end{figure}

\begin{figure}[p]
\noindent\centerline{\epsfig{figure=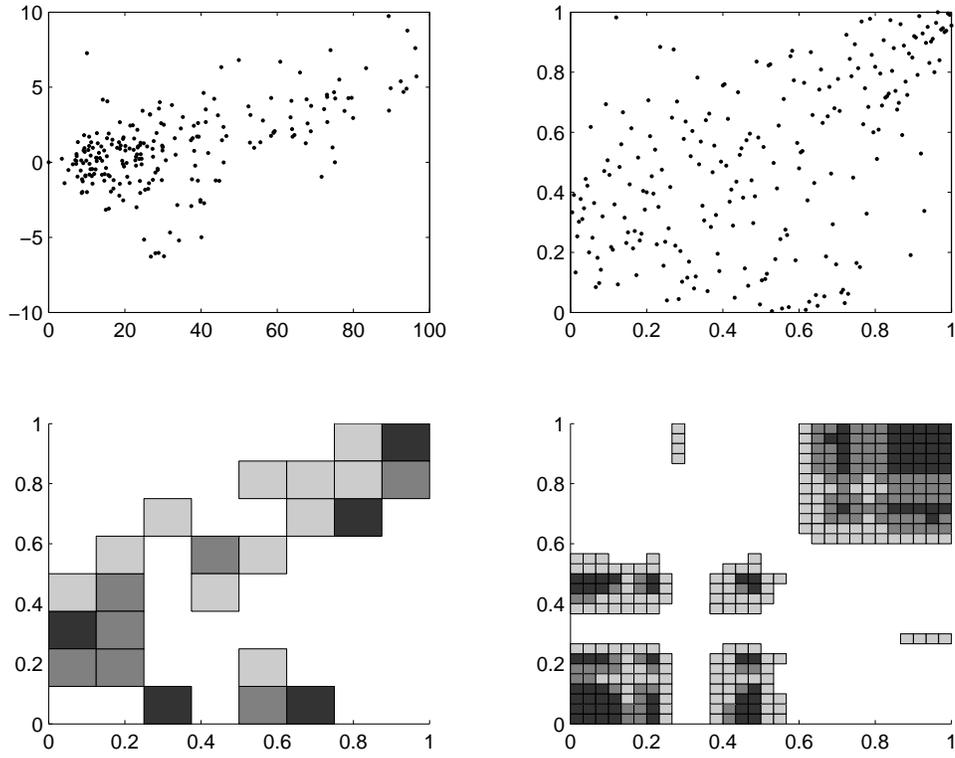,width=0.8\textwidth}}
\caption{Estimation of high probability regions $Q_\alpha$ from real data.  
Dark Grey: $\alpha=0.25$, grey: $\alpha=0.5$, light grey: $\alpha=0.75$.  
Top left: real data, top right: real data after rank transformation, 
bottom left: nonparametric estimate, bottom right: semiparametric estimate.  }

\label{levelreal}
\end{figure}

The dataset consists of $n=225$ countries on which two variables
have been measured: $X$, the life expectancy at birth (years) in 2002
of the total population and $Y$, the difference between the life expectancy at birth
of women and men.
The data is available on the following web-site:
{\tt http://www.odci.gov/cia/publications/factbook/}.
According to the PQD test proposed in \cite{SCI2}, these data are PQD.  
The first step is to compute the rank transformations
$u_i=\mbox{Rank}(x_i)/n$ and $v_i=\mbox{Rank}(y_i)/n$,
see figure~\ref{levelreal}.
Then, the generating function $\psi(x)$ is estimated using
the basis~(\ref{basetrigo}). The optimization procedure
yields an expansion of 
$\hat{\psi}(x)$ with respect to 24 basis functions.  
The function $\hat{\psi}(x)$ is plotted in figure~\ref{phirealdata}.
The estimated Spearman's rho are
$\hat{\rho}_{\mbox{\tiny NP}}=52.4\%$
and $\hat{\rho}_{\mbox{\tiny SP}}=40.7\%$.
These values seem to confirm a moderate positive dependence 
between the two variables. The higher the life expectancy is,
the more important the difference between women and men is. 
Finally, one can estimate the high probability regions. 
The same parameters as in the simulation section are used. The results are
presented in figure~\ref{levelreal}. Of course, since the true
function $\psi(x)$ is not known (and perhaps does not exist), it is
only possible to compare the estimations obtained with the nonparametric
and semiparametric estimates $\widehat{P}_{\mbox{\tiny NP}}$
and $\widehat{P}_{\mbox{\tiny SP}}$. The semiparametric estimation reveals two
main groups of countries.  In the first one, both men and women share a
small life expectancy with weak differences between sexes.
In the second one,  both men and women share a
large life expectancy but with important differences between sexes.  
The sample size is not large enough to obtain meaningful results
with the nonparametric estimate.  

\section{Further work}
\label{extension}

\noindent We have presented a method for estimating copulas 
in the bivariate family of copulas $C_{s,\psi}(u,v)$ in 
the PQD case.  Even though a test \cite{SCI3,SCI2}
has not rejected the PQD assumption,
deciding if the copula model
$C_{s,\psi}(u,v)$ is adapted to a particular dataset is an opened problem. 
It could be possible to build a goodness-of-fit test based on the comparison of
the estimations  $\hat{\rho}_{\mbox{\tiny NP}}$
and $\hat{\rho}_{\mbox{\tiny SP}}$.  
The test would reject the model $C_{s,\psi}(u,v)$ if the difference
$|\hat{\rho}_{\mbox{\tiny NP}}-\hat{\rho}_{\mbox{\tiny SP}}|$
is too large. 
If the PQD assumption is rejected, the proposed estimation method
cannot be used.  To overcome the identifiability problem,
a possible modification of the method would be
to replace the projection step~(\ref{expanphi})
by the selection of the function $\psi$
in a database leading to the best approximation~(\ref{approxpsi}).



\begin{table}[p]
$$
\begin{array}{|c|c|cc|cc|cc|}
\hline
k	& \rho_k & \mbox{mean}(\hat{\rho}_{\mbox{\tiny SP}}) & \mbox{std}(\hat{\rho}_{\mbox{\tiny SP}}) &  \mbox{mean}(\hat{\rho}_{\mbox{\tiny NP}}) & \mbox{std}(\hat{\rho}_{\mbox{\tiny NP}}) & \mbox{mean}(\varepsilon) & \mbox{std}(\varepsilon) \\
        & \times 10^{-2} & \times 10^{-2} &\times 10^{-2} &\times 10^{-2} &\times 10^{-2} &\times 10^{-2} &\times 10^{-2} \\
\hline
1	& 0    & 0.81	& (6.62) & 0.18 & (11.0) & 8.75 & (3.58)\\
2	& 42.5 & 43.0	& (9.91) & 41.2 & (9.11) & 3.67 & (1.26)\\
4	& 66.4 & 65.8	& (7.11) & 64.3 & (6.11) & 3.01 & (1.33))\\
6	& 71.2 & 70.6	& (7.60) & 68.8 & (6.09) & 3.10 & (1.35)\\
8	& 72.8 & 72.1	& (7.68) & 70.2 & (6.05) & 3.10 & (1.17)\\
\hline
\end{array}
$$
\caption{Estimation of the generating function and of the Spearman's Rho
($\rho_k$). The mean value and the standard deviation of the $L_2$ error
$\varepsilon$ as well as of the estimates $\hat{\rho}_{\mbox{\tiny SP}}$ and
$\hat{\rho}_{\mbox{\tiny NP}}$ are evaluated on 100 repetitions.   }
\label{tabphi}
\end{table}

\clearpage


\end{document}